\documentclass[12pt,a4paper]{article}
\usepackage{graphicx}
\usepackage{amsmath}
\usepackage{amssymb}
\newcommand{\GeV}{\,{\rm GeV}}
\newcommand{\I}{{\it I}}

\begin{document}

\title{A Search for Instantons at HERA}

\author{Birger Koblitz\footnote{Max Planck Institut f\"ur Physik, c/o DESY
FH1, Notkestr. 85, 22607 Hamburg, koblitz@mail.desy.de}
for the H1 collaboration\footnote{invited talk
    given at the Ringberg Workshop on HERA Physics on June 19th, 2001}
  }

\maketitle

\begin{abstract} A search for QCD instanton (\I) induced events in
  deep-inelastic scattering (DIS) at HERA is presented in the
  kinematic range of low $x_{\rm Bj}$ and low $Q^2$.  After cutting
  into three characteristic variables for \I-induced events yielding a
  maximum suppression of standard DIS background to the 0.1\% level
  while still preserving 10\% of the I-induced events, 549 data events
  are found while $363^{+22}_{-26}$ (CDM) and $435^{+36}_{-22}$ (MEPS)
  standard DIS events are expected.  More events than expected by the
  standard DIS Monte Carlo models are found in the data. However, the
  systematic uncertainty between the two different models is of the
  order of the expected signal, so that a discovery of instantons can
  not be claimed.  An outlook is given on the prospect to search for
  QCD instanton events using a discriminant based on range searching
  in the kinematical region $Q^2\gtrsim100\GeV^2$ where the \I-theory
  makes safer predictions and the QCD Monte Carlos are expected to
  better describe the inclusive data.
\end{abstract}



\section{Introduction}
Instantons \cite{Belavin,tHooft} are tunnelling phenomena between
topologically different vacuum states present in non abelian gauge
theories which can not be described by perturbation theory. In the
case of QCD, instantons (\I) induce hard processes violating
chirality. Deep-inelastic scattering (DIS) offers a unique opportunity
\cite{RS1} to discover processes induced by QCD instantons because a
sizeable rate is predicted within ``instanton-perturbation theory''
\cite{MRS,RS2,RS3}, and instanton events exhibit a characteristic
final state signature \cite{GIB,CGRS}.

At the HERA machine, which collides $27.5\GeV$ positrons on $820\GeV$
protons, the predicted cross section is large enough to make an
experimental observation possible, although the expected signal is
still small compared to the standard DIS background. Therefore, finding
discriminating observables based on the expected ``fire-ball'' like
topology of \I-induced events producing a large number of hadrons
and combining these variables to a powerful discriminant is the key
ingredient for a successful search analysis.

In contrast to the first pioneering studies which were concentrated on
relatively few variables characterizing \I-induced events
\cite{CK,H1Ins1,H1Ins2}, the analysis by the H1 collaboration
\cite{CKM1,CKM2} which is presented here uses dedicated variables and
optimized cuts.

\begin{figure}[t]
\begin{center}
\includegraphics[width=0.5\textwidth]{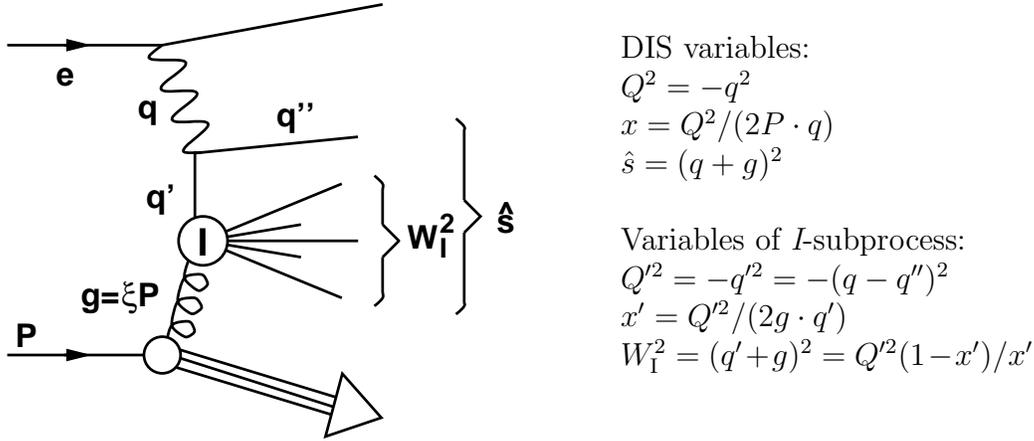}\hfill
\begin{minipage}[b]{.4\textwidth}
DIS variables:\\
$Q^2=-q^2$\\
$x=Q^2/(2P\cdot q)$\\
$\hat s= (q+g)^2$\\
\\
Variables of \I-subprocess:\\
$Q^{\prime 2}=-q^{\prime 2}=-(q-q^{\prime\prime})^2$\\
$x'=Q^{\prime 2}/(2g\cdot q')$\\
$W_{\rm I}^2=(q'+g)^2=Q^{\prime 2}(1-x')/x'$

\vspace{5ex}
\end{minipage}
\caption{
  Sketch of a boson gluon fusion process inducing instanton
  transitions. A virtual photon (with 4-momentum $q$) emitted by the
  incoming electron fluctuates into a $q$-$\bar q$-pair, one of which
  (with 4-momentum $q^\prime$) fuses with a gluon ($g$) out of the
  proton while the other ({\em current quark}) forms a hard jet. The
  \I-subprocess is characterized by the negative 4-momentum squared
  $Q^{\prime 2}=-q^{\prime 2}$ of the in-coming quark and $x'$ which is defined in
  analogy to the standard DIS variable $x_{\rm Bj}$.}
\label{fig:kine}
\end{center}
\end{figure}

\section{Instanton-induced processes in DIS}
Instanton processes in DIS are dominantly induced by photon gluon
fusion as sketched in figure~\ref{fig:kine}, where a photon emitted by
the incoming electron splits into a $q$-$\bar q$-pair. One of these
quarks ($q'$) enters into the \I-subprocess. The basic reaction in this
subprocess is
$$
\gamma^* + g \Rightarrow\sum_{\rm flavours}(\bar q_R+q_R)+n_gg $$
where $q_R$ ($\bar q_R$) denote right handed quarks (anti-quarks) and
$g$ gluons. In every \I-induced event, one quark anti-quark pair of
all $n_f$ kinematically accessible flavours is produced. Charm and
bottom quarks can in principle be produced, but their cross section is
strongly suppressed. Chirality is violated by these events with
$\Delta \chi=2n_f$.  Anti-Instantons also contribute to the cross
section in the same manner, but here only left handed quarks are
in the final state. The quarks are emitted
isotropically together with a mean of
$\left<n_g\right>\approx\mathcal{O}(1/\alpha_s)\approx3$ gluons and
fragment into a densely populated band of hadrons with a width of
about 1.1 units of $\eta$ \footnote{The pseudo-rapidity $\eta$ is
  defined as $\eta=-\log\tan(\theta/2)$, $\theta$ being the polar
  angle and the proton direction being the positive $z$-axis.}. They
are homogeneously distributed in azimuth. This band together with
the relatively hard jet originating from the current quark form the
characteristic final state of \I-induced events. Since in each
\I-event a pair of strange quarks is produced, in addition an increase
in the amount of strange hadrons is expected in the band region.

The actual number of produced hadrons heavily depends on the squared
centre of mass energy $W_I^2=(q'+\xi P)^2$ of the \I-system. It can be
expressed in terms of $Q^{\prime 2}$ and $x'$ by $W_I^2=Q^{\prime 2}(1-x')/x'$, where
$Q^{\prime 2}$ is the negative four momentum vector squared of the quark
entering the \I-subprocess and $x'$ is a variable describing this
process in analogy to the Bjorken scaling variable $x_{\rm Bj}$ in
DIS. The instanton cross section can be safely calculated \cite{RS2}
in ``instanton perturbation theory'' only for large values of $Q^{\prime 2}$
and $x'$. Towards lower values the cross section calculated
perturbatively steeply increases. Matching the instanton size and
distance distributions in the QCD vacuum, which are fundamental properties of
instantons, to high quality lattice calculations allows to derive
lower boundaries $Q^{\prime 2}>113\GeV^2$, $x'>0.35$ above which the
perturbative calculations are applicable.

The most recent calculations of the cross section for \I-induced
events at the HERA collider requires an additional cut on $Q^2\ge
Q^{\prime 2}\approx113\GeV^2$ to further reduce theoretical uncertainties due
to non-planar diagrams, which are not present in the calculation. In
the kinematic region defined by the cuts required by the perturbative
ansatz $x'>0.35$, $Q^2\ge Q^{\prime 2}\ge 113\GeV^2$ and general additional
cuts $x>10^{-3}$ and $0.1<y<0.9$ the cross section at HERA is
$\sigma_{\rm HERA}^{(I)}=29.2^{+9.9}_{-8.1}\,\rm pb$ \cite{RS3,RS4}.
The error given is only reflecting the uncertainties of
$\Lambda_{\overline{\rm MS}}$. The cut $Q^2>113\GeV^2$ has not yet
been included in the experimental analysis.

\begin{figure}[t]
\includegraphics[width=\textwidth]{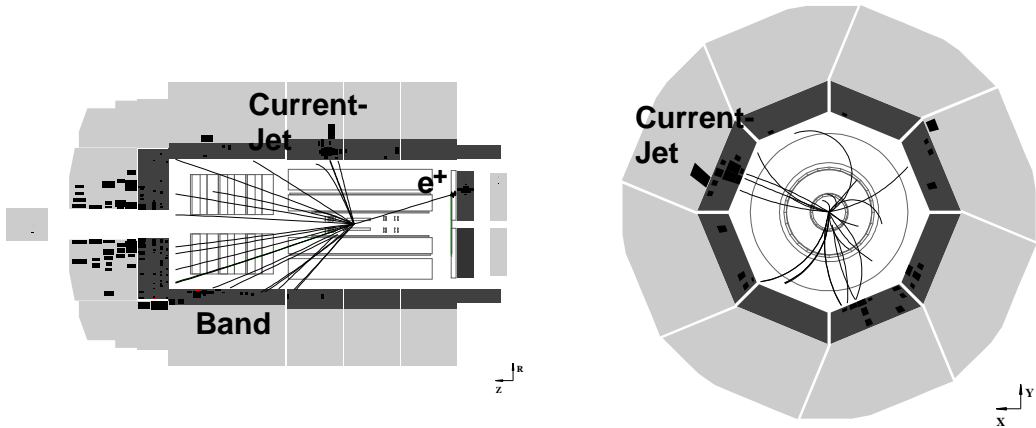}
\caption{Example of a simulated instanton event at 
  $Q^2=54\GeV^2$, $x=2.97\cdot10^{-3}$, $Q^{\prime 2}=136.5\GeV^2$ and
  $x'=0.48$ in the H1 detector. The incoming proton is moving in
  $+z$-direction.}
\label{fig:detector}
\end{figure}

\section{QCD Monte Carlo Generators}
\label{sec:gen}
The instanton induced processes described above are simulated by the
Monte Carlo event generator QCDINS \cite{GRS,RS6}, which works on top
of the HERWIG \cite{Mar} generator and calculates the matrix element
of the hard subprocess. The default implementation of QCDINS is used
with $x'>0.35$, $Q^{\prime 2}>113\GeV^2$. After the hard subprocess has been
generated, gluons are emitted according to the leading logarithm
approximation. The hadronization step describing the transition of the
outgoing partons to hadrons is done using cluster fragmentation
\cite{Web}.

A detailed simulation of the hadronic final state of the background
DIS events is provided by the two Monte Carlo event generators
RAPGAP \cite{Jung} and ARIADNE \cite{Loen}.
Both generators incorporate the $\mathcal{O}(\alpha_s)$ matrix
elements. The main difference is the treatment of gluon emission where
RAPGAP implements the parton shower approach \cite{BenS} while ARIADNE
employs the colour dipole model \cite{GPAL}. We will therefore refer to
RAPGAP as ``MEPS'' and ARIADNE as ``CDM'' in the following.

\section{Event Selection}
\label{sec:detsel}
The data were collected in the year 1997 when
the HERA machine collided $E_e=27.5\GeV$ positrons with $E_p=820\GeV$
protons. The accumulated data sample recorded with the H1 detector
amounted to an integrated luminosity of $15.8\,{\rm pb}^{-1}$. A
detailed description of the H1 detector can be found in \cite{H11}.

A scematic view of a simulated QCDINS event in the H1 detector is given in
figure~\ref{fig:detector}. For this analysis the important detector
components are the backward SpaCal calorimeter which is necessary for
the measurement of the scattered positron and the liquid argon (LAr)
calorimeter which together with the tracking system measures the
particles in the \I-band and the current jet. Because the positron is
only identified using the backward SpaCal calorimeter polar angles are
limited to $156^\circ < \theta_e <176^\circ$ which restricts the
analysis to values of $Q^2<100\GeV^2$.

\section{Finding discriminating variables}
\label{sec:vars}
The most characteristic feature of \I-induced events is the densely
populated band of hadrons in the laboratory frame due to the isotropic
decay of the instanton in its rest system. Therefore the measurement
of the energy flow of the hadronic final state is crucial to the
identification of \I-induced events. The energy flow in the detector
is reconstructed combining the calorimetric energy depositions in
the LAr and SpaCal calorimeters with the momenta of low momentum
tracks ($p_t <2\GeV$) in the central jet chamber according to the
procedure described in \cite{H17}. Also the second striking feature of
\I-induced events, the current jet, relies on this measurement.

From a theoretical point of view the most interesting observables
characteristic to \I-induced events are $x'$ and $Q^{\prime 2}$ since they are
linked to the instanton size and distance which show a characteristic
behaviour.  Therefore reconstructing these kinematic variables of the
\I-subprocess from the hadronic final state is paramount to
identifying \I-induced events.

The starting point of the reconstruction of the underlying kinematics
can either be finding the current jet or reconstructing the expected
band of hadrons in the detector. Both ways have been intensively
studied \cite{Gerigk} and a preference to identifying the band first
was given. The first step is to reconstruct the mean pseudo-rapidity
$\bar\eta$ of the band by finding the $E_t$ weighted mean $\eta$ of
the calorimeter clusters
$$ \bar\eta =\frac{1}{\sum E_t}\sum_{\rm clusters}\eta\,E_t $$
and define the band by $\eta_B\in[\bar\eta-1.1,\bar\eta+1.1]$. This
spread of 1.1 rapidity units of the band is expected from basic
theoretical considerations.

\begin{figure}[t]
\includegraphics[width=\textwidth]{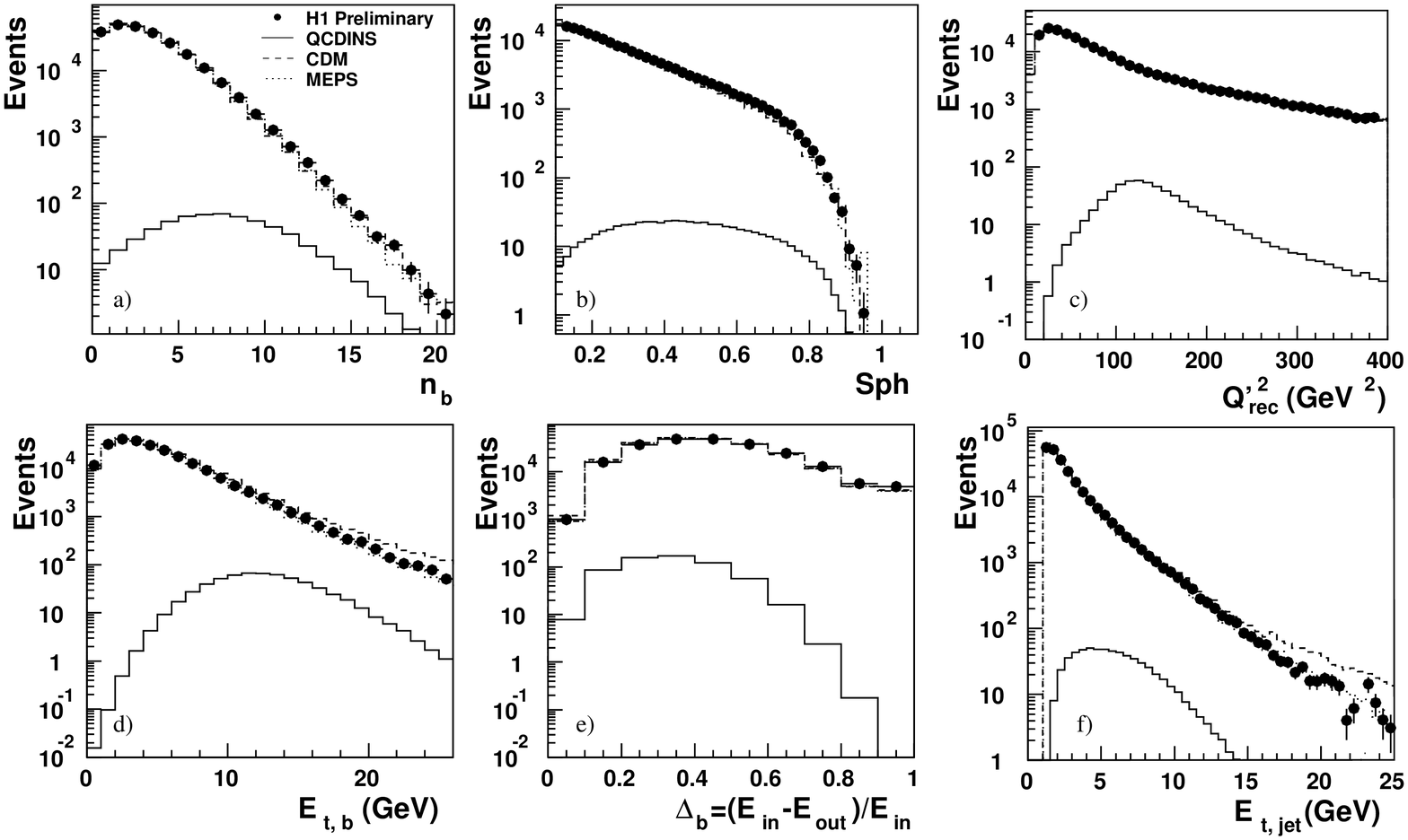}
\caption{
  The distributions of three variables which
  are used to discriminate between standard DIS background and \I-induced
  events are shown in plots a)---c). They are the number of charged
  particles in the \I-band $n_{\rm b}$, the sphericity $\rm Sph$ and
  the reconstructed virtuality of the quark entering the \I-subprocess
  $Q^{\prime 2}_{\rm rec}$. Plots d)---f) show the transverse energy in the
  \I-band $E_{\rm t, b}$, the isotropy variable $\Delta_{\rm b}$ and the
  transverse energy of the supposed current jet.
  }
\label{fig:varslog}
\end{figure}

The second step is to find the current jet which has roughly an $E_t$
of $5\GeV$, as can be seen in figure~\ref{fig:varslog}. Using the
CONE jet algorithm with a cone radius of
$R=\sqrt{(\eta^2+\phi^2)}=0.5$ allows to identify the jet in about
70\% of the cases. The reason for the relatively small cone size is
the fact that the correct current jet can also be directed into the
\I-band and thus a bigger cone radius is prone to pick up more energy
from the band while the cone size is still sufficient to find jets and
reconstruct their kinematics reliably when they are well separated
from the band. If the jet is emitted into the same $\eta$ and $\phi$
region as the band, then the clusters belonging to the jet are removed
from the band.  After the reconstruction of the jet $Q^{\prime 2}$ can be
calculated and assuming the invariant mass of the band to be the mass
$W_I$ of the instanton allows to derive $x'$ (see
figure~\ref{fig:kine}).

The tagged band can now also be exploited to calculate properties that
should be characteristic to the particles produced by an instanton
transition. That is a high isotropy of the outgoing particles, a high
number of charged particles which can be seen in the tracking system
of the detector and possibly an increased strangeness production.

\begin{figure}[t]
\includegraphics[width=\textwidth]{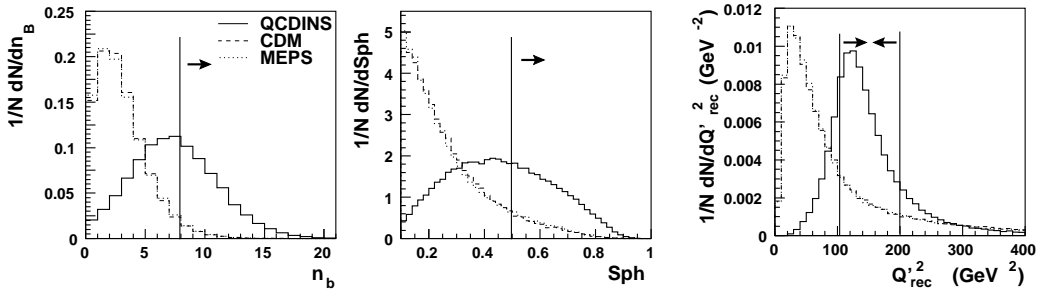}
\caption{ The shape normalized distributions of three variables which
  allow discriminating between standard DIS background and \I-induced
  events. Shown are the number of charged particles in the \I-band
  $n_{\rm b}$, the sphericity $\rm Sph$ and the reconstructed virtuality of
  the quark entering the \I-subprocess $Q^{\prime 2}_{\rm rec}$. The lines and
  the associated arrows denote the cuts applied.}
\label{fig:dvarshape}
\end{figure}

For this analysis the following three variables have been chosen to
characterize the \I-induced events: The reconstructed virtuality of
the quark entering the \I-subprocess $Q^{\prime 2}_{\rm rec}$, the number of
charged particles in the $\eta$-region of the band $n_b$ and the
sphericity of all particles except those belonging to the jet,
calculated in their centre of mass frame. These variables are shown in
figure~\ref{fig:dvarshape}. Because of their potential in
discriminating \I-induced events from background and due to the good
description of the data prior to any cuts these variables are used in
a cut procedure to enhance the \I-signal over the DIS signal.
The distributions are generally described within 5\% by
the standard DIS Monte Carlo programs except for the tails which are
only described within 10\% (for the highest multiplicities up to
20\%).

Three additional observables have been chosen which are used to
monitor the remaining events after the cuts and to see whether they
support the \I-hypothesis after after discriminating against the DIS
background. The variables are the total transverse energy of the band
$E_{\rm t, b}$, the isotropy variable $\Delta_b$ described below and
calculated for the band and the transverse energy of the jet $E_{\rm
  t, jet}$. These variables are described not as well as the first
three, but still the description differs by about only 10\%.

The isotropy variable of the band $\Delta=(E_{\rm in, b}-E_{\rm out,
  b})/E_{\rm in, b}$ is defined using the quantities
$$ E_{\rm out} = \min_{\vec i} 
  \sum_{n\; {\rm Hadr.}}\left|\vec p_n\cdot\vec i\right|
\qquad
E_{\rm in} = \max_{\vec i} 
  \sum_{n\; {\rm Hadr.}}\left|\vec p_n\cdot\vec i\right|
$$
where $E_{\rm out}$ will be small for a non-isotropic set of hadrons since
then there is a direction $\vec i$ onto which the projections of the
hadron momenta $\vec p_n$ is small and $E_{\rm in}$ will be large for
the same set of non-isotropically distributed hadrons. Therefore
$\Delta$ is approximately unity for a non-isotropic distribution and
vanishes for isotropic distributions.

\section{Search Strategy and Results}
\label{sec:loqres}
In order to reduce the standard DIS background, several different
combinations of cuts into the three observables $Q^{\prime 2}_{\rm rec}$, $\rm
Sph$ and $n_b$ are studied and evaluated for their ability to
separate \I-induced events from the background by achieving a high
separation power $S=\epsilon_{\rm INS}/\epsilon_{\rm DIS}$, where
$\epsilon$ is the fraction of events left after the cuts. 125 different
scenarios where surveyed which represent all possible combinations of
five different values at which to cut into the three observables. The
following cuts are used: $n_b>5,6,7,8,9$, ${\rm Sph}>0.4, 0.5, 0.55,
0.6, 0.65$ and for $Q^{\prime 2}_{\rm rec}$: $95, 100, 105, 110, 115 <
Q^{\prime 2}_{\rm rec}<200\GeV^2$. Three different scenarios were singled out
according to the following criteria:
\begin{description}
\item[(A)] The highest instanton efficiency ($\epsilon_{\rm INS}$)
\item[(B)] High $\epsilon_{\rm INS}$ at reasonable background
  reduction ($\epsilon_{\rm DIS}$)
\item[(C)] Highest separation power $S$ at $\epsilon_{\rm INS}\approx
  10\%$
\end{description}

The values of the cuts chosen in the three scenarios as well as the
number of events in the data and the two standard DIS Monte Carlo
simulators is shown in table~\ref{tab:results}. The error on the
number of MC events includes systematical and statistical
uncertainties.  The systematic experimental uncertainties are of an
order of 5--10\% and the two available standard DIS Monte Carlo models
based on leading order QCD matrix elements and implementing different
treatments of higher order parton emissions agree with each other
within 20\%.  The number of data events is about 20\% larger than
predicted by the DIS models. Note that this is about the difference of
the prediction of the two Monte Carlo generators and the background
estimation is largely unknown in this extreme phase space region.
Whether the excess can be explained by instanton induced processes or
is simply due to an incomplete implementation of higher order parton
emissions or due to another possible deficiency in the description of
the DIS background, for example, an incorrect treatment of heavy
quarks, remains an open question.

\begin{figure}[t]
\includegraphics[width=\textwidth]{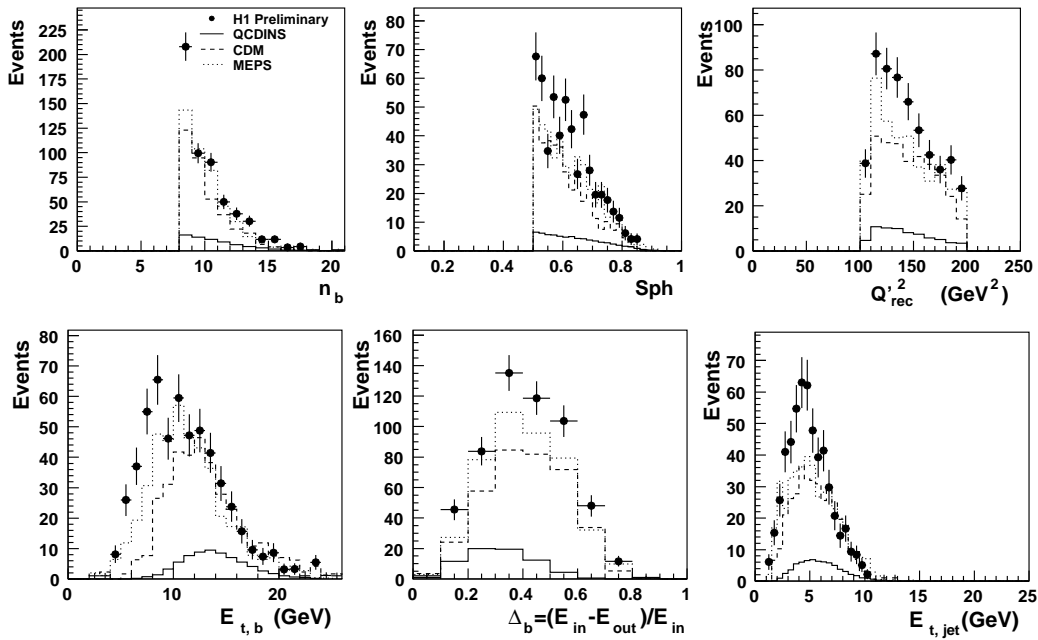}
\caption{The six observables after the cuts to enhance the instanton
  signal of cut scenario C.
}
\label{fig:varscut}
\end{figure}

Looking at the three variables after the cuts of scenario~C
(figure~\ref{fig:varscut}) shows that the
three variables which were used in the cut scenario would support the
\I-hypothesis while especially the distribution of $E_{\rm t, b}$ and
$E_{\rm t, Jet}$ disfavour it. In both of these distributions data
shows a softer behaviour than the DIS Monte Carlo programs while
QCDINS predicts a harder $E_{\rm t}$ spectrum.  Recent theoretical
analysis shows, that this might be due to the lack of a cut in
$Q^2>113\GeV^2$ which suppresses the contribution of non-planar graphs
in the \I-subprocess and which are not implemented in QCDINS. In fact,
Schrempp and Ringwald have shown that it is possible to soften the
$E_{\rm t}$ spectra of the jet and the band while not affecting the
other four distributions by a cut against low $x$ and thus, by
exploiting the strong correlation between $x$ and $Q^2$ in QCDINS and
suppressing the contribution of events at low $Q^2$ \cite{RS7,SRingberg}.

It will therefore be definitely interesting to extend the analysis to
high $Q^2$ where the theoretical uncertainties in particular on these
two variables are smaller. In the following
section a possible search strategy is presented.

\begin{table}
\caption{\label{tab:results} The three cuts scenarios A, B and C which
  are used to reduce the background. }
\begin{tabular}{|l|c|c|c|c|c|c|c|c|c|}
  \hline
  \multicolumn{4}{|l|}{Scenario\hspace{2cm}Cuts} & $\epsilon_{\rm INS}$ &
  \multicolumn{2}{|c|}{$\frac{\epsilon_{\rm INS}}{\epsilon_{\rm DIS}}$} &
  \multicolumn{3}{|c|}{\# Events}\\
  & $Q^{\prime 2}\;[\GeV^2] $& $Sph >$& $n_{\rm b}\ge$& &CDM & MEPS & CDM &
  MEPS & Data\\
  \hline
  A& 95--200 & 0.4 & 5 & 32\% & 35 & 34 & $2469^{+242}_{-238}$ &
  $2572^{+237}_{-222}$ & 3000 \\
  B& 105 -- 200 & 0.4 & 7 & 21\% & 56 & 52 & $1005^{+82}_{-70}$ &
  $1084^{+75}_{-46}$ & 1332 \\
  C & 105 -- 200 & 0.5 & 8 & 11\% & 86 & 71 & $363^{+22}_{-26}$ & 
  $435^{+36}_{-22}$ & 549 \\
  \hline
\end{tabular}
\end{table}

\section{A Prospect of a Search at $Q^2>113\GeV^2$}
\label{sec:rsres}
Since the expected cross section of \I-induced events in the kinematic
region ($x_B>10^{-3},\,0.1<y<0.9,\, Q^2>113\GeV^2$) is by
approximately two orders of magnitude lower than the DIS cross section
$\sigma_{\rm DIS}\approx 3000\,\rm pb$, it is important to gain the
highest possible signal to background ratio. This can be achieved by
exploiting observables characterising \I-induced processes. To find
these observables a large number of promising event variables have to
be investigated and the sensitivity to systematic details in the
modelling of the hadronic final state has to be tested.  This requires
a sophisticated and fast discrimination method to find the appropriate
combination of event variables. Range Searching allows to define such
a powerful discriminant.

\subsection{A Discriminant Based on Range Searching}
\begin{figure}[t]
\begin{center}
\includegraphics[width=0.6\textwidth]{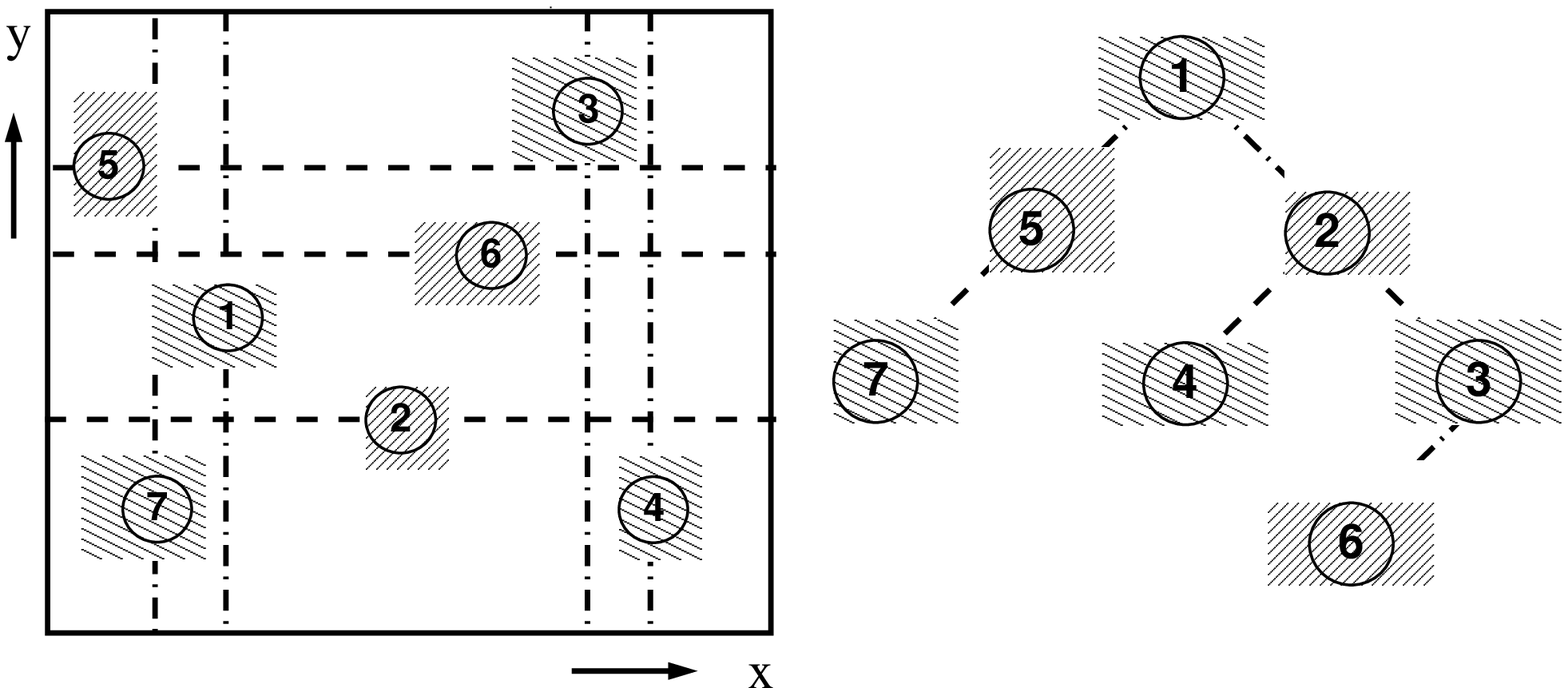}
\caption{Example of a binary tree storing events 1--7 according to
  their position in the $x$-$y$ plane. The first event is always the
  top node of the tree which has two daughter nodes for events with
  smaller and larger $x$-value. On the second level the tree then
  sorts according to $y$.}
\label{fig:btree}
\end{center}
\end{figure}

Events can be classified as signal or background by estimating their
probability density $\rho$ at the point of the event in the
event-variable phase space, employing a Monte Carlo (MC) generator to
sample the densities. In the case of neural networks (NN) this is done
by fitting the probability-density with the adjusted weights of the
neurons. To circumvent this time consuming procedure the density at
each point can be directly estimated by counting the number of
background and signal events in a surrounding box $V$. Given the ratio
$$\ell:=\frac{\rho(I)}{\rho(DIS)}=\frac{\#I(V)}{\#DIS(V)}$$
the probability of an event to be
a signal event is $D=\ell/(1+\ell)$. Compared to NN's this method also
has the advantage of not extrapolating into phase space regions where
there are no sample events available. Thus signals from data
events outside the region covered by the MC simulation can be
avoided. Counting the number of
events in the vicinity of a certain point is a problem known as Range
Searching.

Range Searching algorithms have been developed which allow a search
time $\sim \log(n)$, where $n$ is the number of points that have to be
searched \cite{Sedg}. An algorithm using binary trees to store the
events is employed, suitable especially for a large number of events
and dimensions (i.e.  observables).  The MC events are successively
filled into the nodes of two binary trees, one for the signal events
and one for the background events, where the criterion by which the
decision is taken to descend to the left or right of a node is given
by the value of one of the event variables (see
figure~\ref{fig:btree}).  While descending the tree this variable
cycles through all considered discriminating variables.  After
filling, the position of every event in the tree is given by its
coordinates in the event variable space.  Classification of an event
is done by searching in the trees for all background and signal events
in the box $V$. This is done in the same manner as filling the tree. A
more detailed description and the properties of this method can be
found in \cite{BT1}.

\begin{figure}[t]
\includegraphics[width=\textwidth]{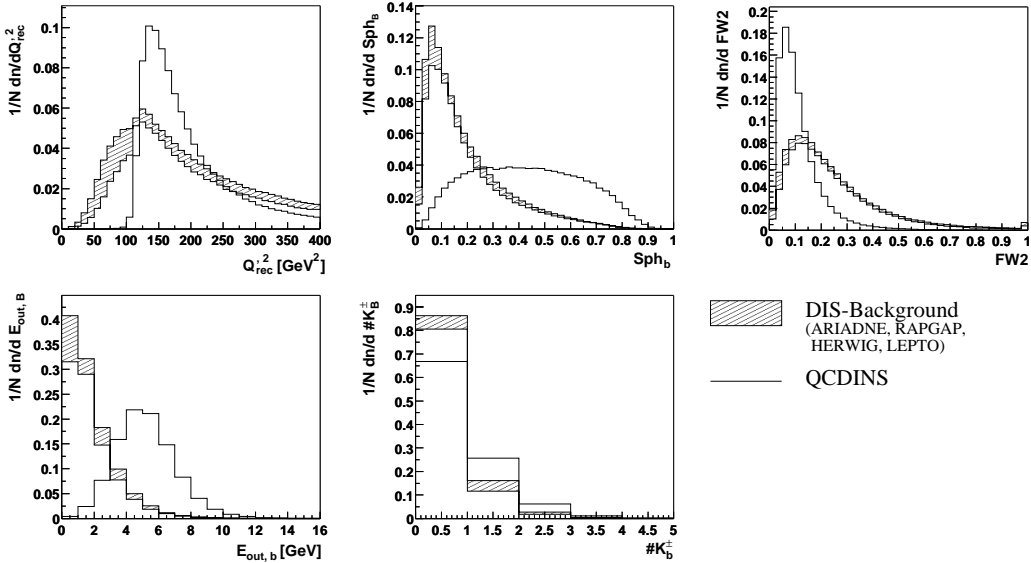}
\caption{The characteristic event variables providing good
  instanton separation with small systematic uncertainties. Shown is
  the reconstructed virtuality of the quark entering the \I-subprocess
  $Q_{\rm rec}^{\prime 2}$, the sphericity of the particles not associated to
  the jet in their rest system, the second Fox-Wolfram moment of these
  particles and the event shape variable $E_{\rm out, B}$ which is the
  projection of the particle transverse energy onto the axis, that
  makes this quantity maximal (see \cite{CGRS}). Finally the number of
  charged kaons in the \I-Band is shown.  }
\label{fig:hq2vars}
\end{figure}

\subsection{Results}
\begin{figure}[t]
\begin{center}
\includegraphics[width=0.9\textwidth]{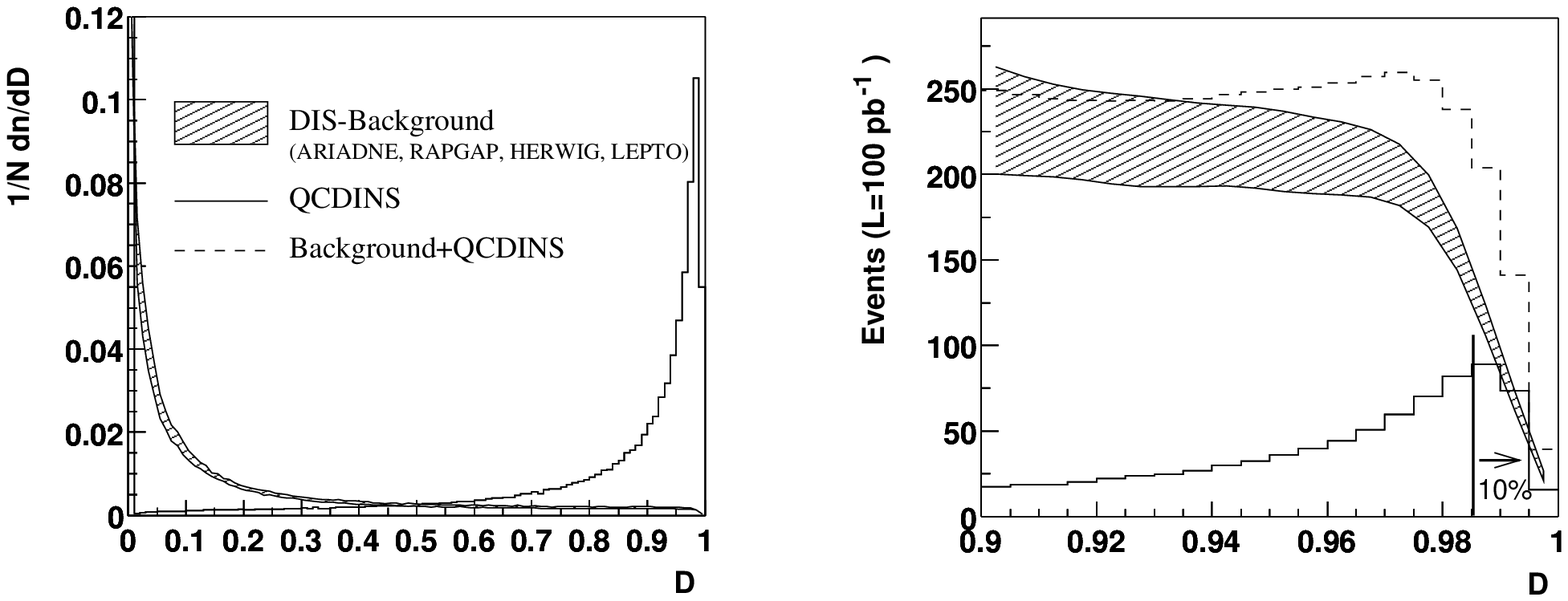}
\caption{To the left, the shape normalized discriminant $D$ for the
  instanton events using QCDINS and for standard DIS events using four
  MC simulators is shown. The second plot shows a zoom into the rightmost
  part and is now normalized to a luminosity of $L=100\,{\rm
    pb^{-1}}$. At $\epsilon_{\rm INS}=10\%$ 178 \I-induced events are
  expected.}
\label{fig:hq2res}
\end{center}
\end{figure}

Starting with 35 variables based on the hadronic final state the best
12 were chosen by calculating the discriminant with all 2-combinations
(pairs) of the initial variables and taking those variables which
provide a high separation power $S=\epsilon_{\rm
  INS}/\epsilon_{\rm DIS}$ demanding an efficiency for instantons of
$\epsilon_{\rm INS}=10\%$. The number of considered variables is
further reduced by calculating all 5-combinations and selecting those
with the highest separation power and a small systematic variation of the
background.  The systematic uncertainty was obtained by using four
standard DIS-MC simulators which were tuned to data on representative
hadronic final state quantities, in the range $Q^2>100\GeV^2$ at HERA
\cite{NBrook}. The variables forming the best combination is shown in
Figure~\ref{fig:hq2vars}.  The separation power for $\epsilon_{\rm
  INS}=10\%$ is $S=126$. In Figure~\ref{fig:hq2res} the shape
normalized discriminant $D$ is shown for the \I-induced and the
background events, as well as the distribution for $D>0.9$ normalized
to a luminosity of $100\,{\rm pb^{-1}}$ which is comparable to that
already collected by each of the HERA experiments H1 and ZEUS. An
event sample can be isolated where half of the events are expected to
be due to instantons while the \I-efficiency is still 10\%.

\begin{figure}[t]
\begin{center}
\includegraphics[width=0.85\textwidth]{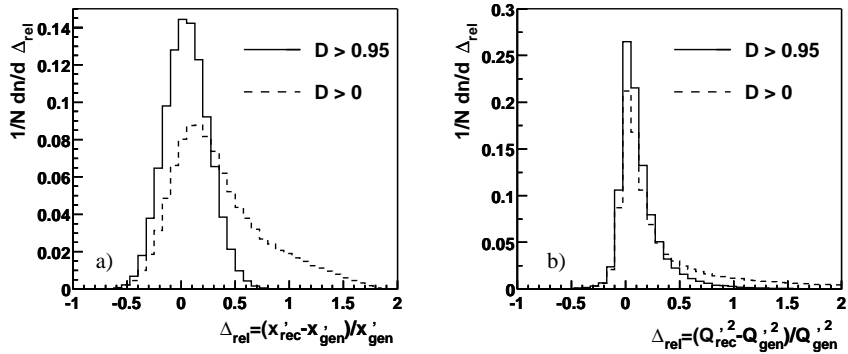}
\caption{The reconstruction of $x'$ and $Q^{\prime 2}$ strongly improves when
  a cut on the discriminator is applied. With a cut $D>0.95$ the RMS
  of the reconstruction of $x'$ is 0.20 and 0.17 for $Q^{\prime 2}$.
}
\label{fig:rec}
\end{center}
\end{figure}

In addition the reconstruction of $x'$ and $Q'2$ was studied before
and after a cut on $D$. As can be seen from figure~\ref{fig:rec}, a cut
on the discriminant at $D>0.95$ improves the reconstruction of both
variables strongly. The events which were selected by the
discriminant are very \I-like and therefore the reconstruction based
on ideal \I-events is successful.

\section{An Outlook on Proving Chirality Violation}
The one property of \I-induced events which is not present in standard
deep inelastic scattering is chirality violation. Therefore finding
events which violate chirality would be a strong hint for \I-induced
processes.

In every \I-induced event chirality is violated by $\Delta\chi =2n_f$,
meaning that all the outgoing quarks are right handed (or all left
handed, in the case of anti-instantons). In addition, in every
\I-induced event a pair of $s$-$\bar s$ quarks is produced. The decay
of $\Lambda$ baryons might provide a unique way of measuring the
polarization of these strange quarks, provided both quarks form a
$\Lambda$ and $\bar\Lambda$-baryons after hadronization respectively.
However, the problem here is that a large fraction of the
$\Lambda$-baryons contain $s$-quarks stemming from fragmentation.  So
far, no efficient selection mechanism has been found to select
$\Lambda$-pairs containing only strange quarks produced in the hard
instanton process. Inferring the polarization of the $s$-quark from
the $\Lambda$-decay has been well studied in polarized deep inelastic
scattering \cite{MSSY}.

An additional problem of such an analysis is the relatively low
detector efficiency for $\Lambda$-baryons which is of the order of
25\% for a rapidity range of $|\eta|<1.3$ for the H1-detector, which
leads to only a 6\% chance of finding both $\Lambda$-baryons in an
event.

\section{Conclusions}
H1 has conducted a search for \I-induced events in the kinematic range
of $x_{\rm Bj} > 10^{-3}$, $0.1<y<0.6$ and $\theta_{\rm el} >
156^\circ$ using 1997 data. Using a simple cut technique the standard
DIS background has been suppressed up to the 0.1\% level while
retaining 10\% of the \I-induced events. More events were found in the
data than predicted by the DIS Monte Carlo programs.  Nevertheless
this excess is not conclusive since it is of the order of the
difference of the Monte Carlo predictions. While four out of the six
distributions that were studied would support the \I-hypothesis after
these cuts, the $E_{\rm t}$ spectra of the band and the jet do not
favour it. However, recent theoretical analysis has shown that this
discrepancy can be explained by the lack of a cut $Q^2>113\GeV^2$
which suppresses non-planar graphs not present in the QCDINS Monte
Carlo simulator.

It is therefore of high interest to extend the analysis to this range
of $Q^2$ to reduce the theoretical uncertainties. A first survey of
the prospects of such a search was done using the Range Searching
method to discriminate between the \I-signal and the DIS
background. The results are very promising since this method allows to
choose combinations of observables with small systematic
uncertainties in the final discriminant and a high separation power.

A possibility of measuring the chirality violation of the \I-induced
events via the decay of $\Lambda$-baryons is in principle possible
if an efficient mechanism to select $\Lambda$-baryons containing only
strange quarks from the hard \I-process can be found and experimental
difficulties can be overcome. In any case such an analysis will
require a very large luminosity and an isolated data sample where the
contamination from standard DIS processes is low.

\section{Acknowledgements}
I would like to thank A. Ringwald and F. Schrempp for many enlightening
and delighting discussions about instantons and for their invaluable
tool QCDINS. T. Carli and S. Mikocki I would like to thank for their
fruitful collaboration in this hunt for instantons. I am very
grateful to Chr. Risler for providing the information on
$\Lambda$-identification, and M. zur Nedden and P. Schleper for their
careful reading of this paper.

\end{document}